# Maximizing spin-orbit torque efficiency of Ta(O)/Py via modulating oxygen-induced interface orbital hybridization

Liupeng Yang[1], Yining Fei[1], Kaiyuan Zhou[1], Lina Chen[2, *], Qingwei Fu[1], Liyuan Li[1], Cunjie Yan[1], Haotian Li[1], Youwei Du[1], and Ronghua Liu[1,*]

[1]*National Laboratory of Solid State Microstructures, School of Physics and Collaborative Innovation, Center of Advanced Microstructures, Nanjing University, Nanjing, 210093, China*
[2]*New Energy Technology Engineering Laboratory of Jiangsu Provence & School of Science, Nanjing University of Posts and Telecommunications, Nanjing 210023, China*

Spin-orbit torques due to interfacial Rashba and spin Hall effects have been widely considered as a potentially more efficient approach than the conventional spin-transfer torque to control the magnetization of ferromagnets. We report a comprehensive study of spin-orbit torque efficiency in Ta(O)/Ni$_{81}$Fe$_{19}$ bilayers by tuning low-oxidation of β-phase tantalum, and find that the spin Hall angle $\theta_{DL}$ increases from ~ -0.18 of the pure Ta/Py to the maximum value ~ -0.30 of Ta(O)/Py with 7.8% oxidation. Furthermore, we distinguish the efficiency of the spin-orbit torque generated by the bulk spin Hall effect and by interfacial Rashba effect, respectively, via a series of Py/Cu(0-2 nm)/Ta(O) control experiments. The latter has more than twofold enhancement, and even more significant than that of the former at the optimum oxidation level. Our results indicate that 65% enhancement of the efficiency should be related to the modulation of the interfacial Rashba-like spin-orbit torque due to oxygen-induced orbital hybridization cross the interface. Our results suggest that the modulation of interfacial coupling via oxygen-induced orbital hybridization can be an alternative method to boost the change-spin conversion rate.

Since the achievement of current-induced spin-transfer torque (STT) driving magnetization switching and auto-oscillation in magnetic multilayer-based nanodevices for the first time in the late 1990s, all-electric control of magnetization reversal has been receiving massive studies and continuous interest for realizing high-performance, low-power spintronic memory, and logic devices.[1-3] Very recently, there is another alternative approach to generate spin-torque in the magnetic heterostructures consisted of nonmagnetic heavy metals (NM), topological materials and ferromagnetic materials (FM) based on the spin-orbit interaction (SOI) mechanism.[4,5] The pure spin current generated by spin Hall effects in the materials with strong spin-orbit coupling (SOC) can exert spin torque on the magnetization of the free layer.[6,7] Except for the spin Hall effect (SHE), the interfacial Rashba-Edelstein effect (IREE) due to the broken inversion symmetry of heterostructure also allows charge-spin conversion with an efficiency comparable to or larger than that of SHE, especially in the ultrathin NM/FM heterostructure or multilayer systems.[8-10]

The parameter describing the charge-to-spin current conversion rate is the spin Hall angle $\theta_{SH}$, which is defined as the ratio of spin and charge current density ($J_S/J_C$).[4,11] The parameter $\theta_{SH}$ is usually considered the indicator of spin-orbit torque (SOT) efficiency, which governs fundamental device characteristics such as energy consumption and operation speed.[12,13] Therefore, higher SOT efficiency is highly desirable for developing spintronic devices. The $\beta$-phase tantalum ($\beta$-Ta) and tungsten ($\beta$-W) show a larger bulk spin Hall angle $\theta_{SH}$ ~ -0.15 and -0.30, respectively.[14,15] In NM/FM with broken inversion symmetry, except SHE in the bulk NM, the additional IREE also can be utilized to boost current-induced SOT efficiency further.[11,16] The SOT efficiency arising from interfacial Rashba SOI is determined by Rashba coefficient $\alpha_R$, expressed by the Hamiltonian: $H_R = \alpha_R \cdot (\hat{z} \times k)$, where $k$ is momentum, and $\hat{z}$ is the unit normal to surface or interface.[17-19] It can be modulated by tunning chemical composition, strain and applying the electric field perpendicular to the interface.[20-22] It was recently reported that the charge-to-spin current conversion rate arising from SOT could be further enhanced by oxygen incorporation into heavy metal W and Pt film.[23,24] However, it still lacks sufficient detail about the working condition parameter optimization for the charge-spin conversion rate and the behind physical mechanism in the FM/NM system.

Here, we systematically studied that the charge-spin conversion rate of the Ta(O)/Py system and found the SOT efficiency has 65 % enhancement at a 7.8 % oxidation level. To get insight into the origin of enhancement, we performed a series of control experiments on Ta(O)/Cu(0-2 nm)/Py. We found that partial oxidation induced SOT enhancement behavior in Ta(O)/Py completely disappears after inserting only 1 nm Cu between Ta(O) and Py layers, indicating that SOT enhancement should originate from the modulation of the interfacial Rashba-like spin-orbit torque due to oxygen-induced orbital hybridization in the interface between Ta(O) and Py. Furthermore, SOT efficiency contributed from the bulk spin Hall effect and the interfacial Rashba effect is also distinguished by analyzing Cu thickness-dependent spin Hall angle of Ta(O)/Cu/Py. Our results demonstrate that the low oxidation of Ta can dramatically enhance the interfacial Rashba SOT to more than twofold, while it has a negligible impact on the bulk SHE and electrical conductivity of the Ta layer.

The films of Ta(O)(8)/Py(6), with film thickness in nm in parentheses, were deposited by DC magnetron sputtering on annealed sapphire substrates with (0001) orientation. The base pressure before the deposition was less than $3\times10^{-8}$ Torr, and the pressure during deposition was 3.7 mTorr. The Ta(O) layer was grown first using the mixture gas of $O_2$ and Ar with a gas flow ratio of $O_2$ and Ar varying from 0 to 0.53%. After Ta(O) layer growth, the sputtering chamber was pumped to less than $3\times10^{-8}$ Torr base pressure again before the deposition of the Py layer. All films were covered in-situ by 2 nm MgO capping layer to prevent Py from exposure to the atmosphere. For the fabrication of devices used in the ST-FMR experiment, the films were patterned into $5\times10$ μm$^2$ rectangular shape by combining electron beam lithography (EBL) and ion milling. The oxygen concentration of Ta(O) was determined by X-ray photoelectron spectroscopy (XPS).

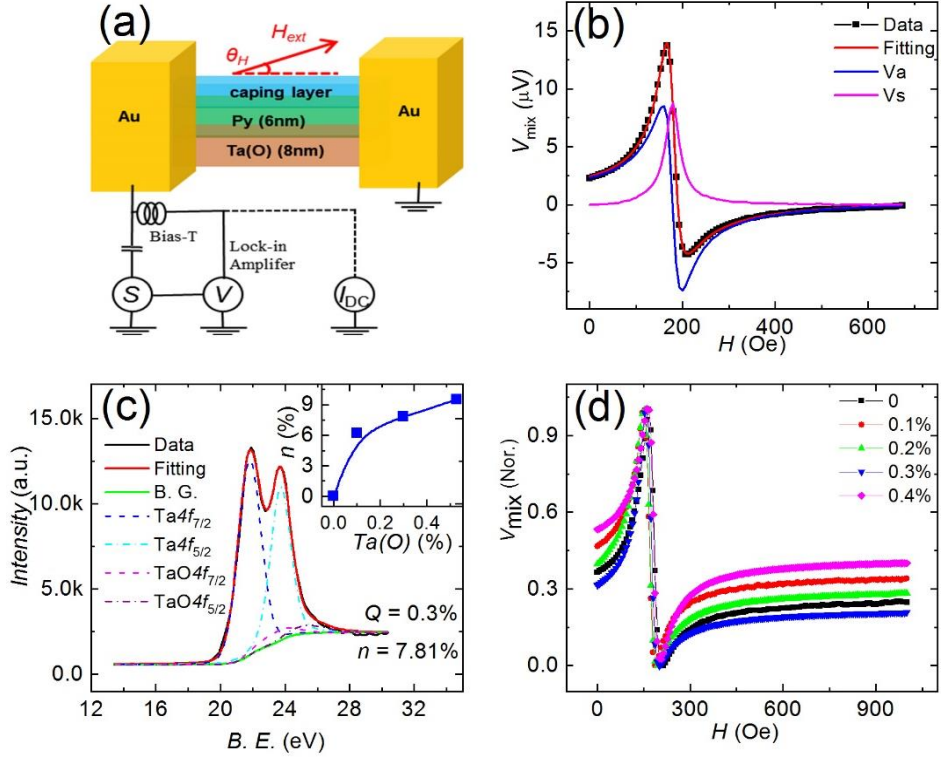

Fig. 1(a) Schematics of ST-FMR measurement setup for Ta(O)(8)/Py(6), with film thickness in parentheses in nm. (b) Typical experimental data and fitting curves of field-dependent $V_{\text{mix}}$ at an excitation frequency $f = 4$ GHz. $V_s$ and $V_a$ correspond to the symmetric and anti-symmetric Lorentzian components, respectively. (c) XPS spectra of Ta(O)(8) deposited at the oxygen flow rate $Q = 0.3\%$, blue and cyan curves are Ta $4f_{7/2}$ and $4f_{5/2}$ peaks, and magenta and purple curves are TaO $4f_{7/2}$ and $4f_{5/2}$ peaks. The inset shows oxygen concentration $n$ versus oxygen flow rate $Q$. (d) Normalization of field-dependent $V_{\text{mix}}$ spectra obtained at $f = 4$ GHz for the five different samples grown at marked oxygen flow rate $Q$.

The spin-orbit torques efficiency of Ta(O)/Py bilayers is determined firstly by the spin-torque ferromagnetic resonance (ST-FMR) technique.[4] Fig. 1(a) shows the schematic of the ST-FMR measurement setup. In our ST-FMR experiments, a radio-frequency current ($I_{\text{rf}}$) passes through the device along its longitudinal direction, and an in-plane external magnetic field $H$ forms an angle of 30° with the longitudinal direction. The $I_{\text{rf}}$ in the Ta(O) layer generates oscillating SOTs (damping-like torque and field-like torque) and the Oersted field, which acts on the magnetic Py layer and drive the magnetization oscillation around the external magnetic field $H$. The magnetization precession of the Py layer results in an oscillation resistance due to the anisotropic magnetoresistance (AMR) effect. Consequently, the oscillating resistance and $I_{rf}$ produce a dc mixing voltage $V_{mix}$ across the device, detected by a Lock-in Amplifier. Besides, the magnetization oscillation also pumps spin current into the adjacent Ta layer through the interface. It generates a transversal dc voltage across the device due to the inversed spin Hall effect of the bulk Ta and interfacial Rashba-type spin-to-charge conversion effect. Based on both spin-to-charge conversion effect and spin rectification of magnetoresistance, the ST-FMR technique has been widely used to study dynamics and spin-dependent transport of the micro-sized or even nano-sized spintronic devices.[25-27]

The ST-FMR spectra of Ta(O)(8)/Py(6)/MgO(2) were performed by varying the excitation frequency from 3 to 12 GHz. The measured ST-FMR signals $V_{mix}$ can be expressed as

$$V_{mix} = S \frac{w^2}{(\mu_0 H - \mu_0 H_{FMR})^2 + w^2} + A \frac{w(\mu_0 H - \mu_0 H_{FMR})}{(\mu_0 H - \mu_0 H_{FMR})^2 + w^2}, \quad (1)$$

where $S$, $A$, $w$, and $\mu_0 H_{FMR}$ are the magnitude of the symmetric component, the anti-symmetric component, the linewidth, and the FMR field, respectively.[23] Here, the symmetric component is proportional to the damping-like torque, while the anti-symmetric component provides information about the field-like torque and Oersted field torque. Fig. 1(b) shows the representative $V_{mix}$ of Ta(O)(8)/Py(6) ($Q = 0.3\%$) at $f = 4$ GHz and fitting curves. The symmetric components ($V_s$) and anti-symmetric components ($V_a$) of the $V_{mix}$ can be extracted by fitting the experimental data using the Eq. (1).

To quantitively obtain the oxidation level of the bottom Ta(O) layer, we performed X-ray photoelectron spectroscopy (XPS) of Ta(O)(8) layers deposited at different oxygen flow rates $Q$. Based on the XPS data, we found that the Ta(O) layer only has a small proportion of the metastable orthorhombic TaO phase of tantalum oxides, and the oxygen concentration increases from 0 to 9.52% when the value of $Q$ increases from 0 to 0.53%. As shown in Fig. 1(c), for $Q = 0.3$%, the oxygen concentration $n$ of Ta(O) is 7.81%. Fig. 1(d) shows normalization of the mixing voltage $V_{mix}$ spectra of samples deposited at five different oxygen flow rates $Q$ from 0 to 0.4%. To directly compare their lineshapes, we normalized the obtained voltage spectra of the five different samples. One can easily see that the lineshape of $V_{mix}$ spectra changes significantly with the increased oxygen flow rate $Q$.

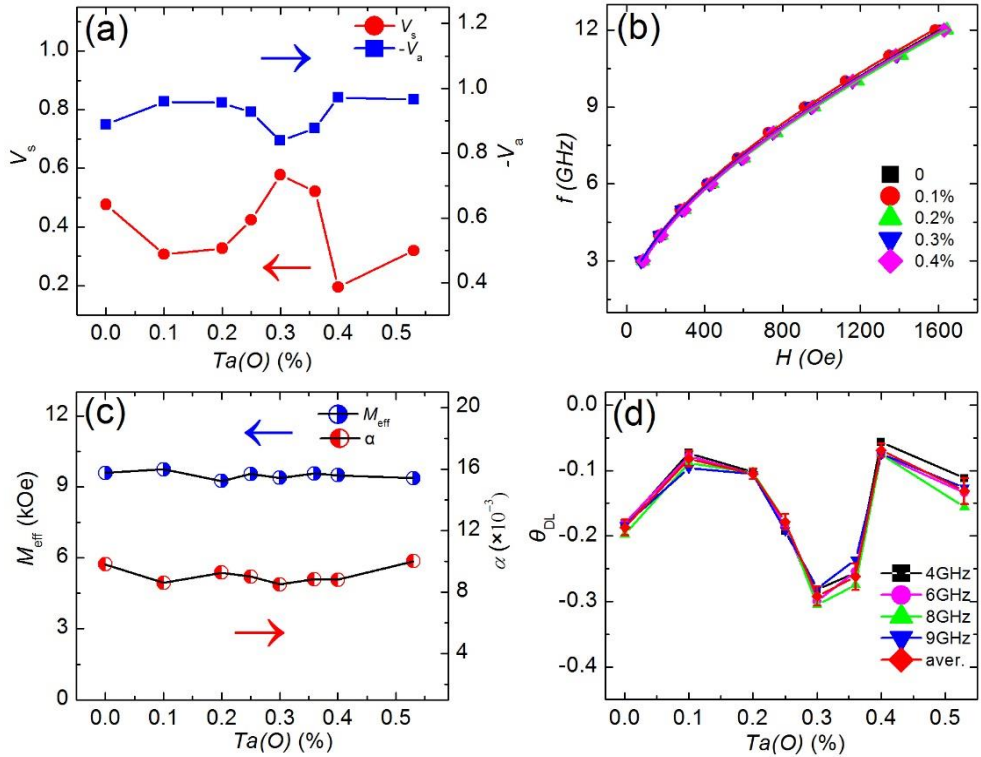

Fig. 2 (a) The symmetric component $V_s$ and the anti-symmetric component $V_a$ of the ST-FMR signal $V_{mix}$ of Ta(O)(8)/Py(6) samples with different oxygen concentrations extracted by fitting the experimental data using Eq.(1). (b) Resonant field $H_{res}$ vs. excitation frequency $f_{ex}$ data (symbols) of the marked samples. The solid curves are the Kittel formula fitting. (c) The effective demagnetization field $M_{eff}$ and damping constant $\alpha$ of the samples. (d) The calculated effective spin Hall angle $\theta_{DL}$ of the studied samples at several different excitation frequencies.

Fig. 2(a) shows the $V_s$ and $-V_a$ extracted from the data of Fig. 1(d). $V_s$ exhibits a non-monotonic dependency on oxygen concentration and has a maximum value of 0.58 around $Q = 0.3\%$, while $-V_a$ exhibits a tendency opposite to $V_s$, and has a minimum value of 0.83 at $Q = 0.3\%$. Fig. 2(b) shows the resonant field $H_{res}$ as a function of the excitation frequency $f_{ex}$ of five samples with different oxygen flow rate $Q$. Fig. 2(c) shows the demagnetization filed $4\pi M_{eff}$ determined from the Kittel formula, $f = (\gamma\mu_0/2\pi)$ $[H_{FMR} (H_{FMR}+4\pi M_{eff})]^{1/2}$, where $\gamma/2\pi$ is the gyromagnetic ratio. One can see that all five Ta(O)(8)/Py(6) samples have the almost same value of $4\pi M_{eff} \sim 9.5$ kOe. The dependence of linewidth $\Delta H$ on the resonant frequency allows us to get the Gilbert damping $\alpha$ by a linear fitting using $\Delta H = (2\pi/\gamma)\alpha f + \Delta H_0$, where $\Delta H_0$ is the inhomogeneous line broadening. Unlike the charge-spin conversion tendency, the Gilbert damping $\alpha$ remains a relatively small deviation around average $\sim 0.01$ for different oxygen flow rates $Q$. The reason is the damping depends not only on the spin currents injected into the ferromagnetic layer but also on the ratio of the elastic scattering time to the spin-flip scattering time, as well as the spin mixing conductance.[14,28]

Based on the symmetry analysis of spin torques and many previous reports of the NM/FM bilayer system with strong spin Hall effect,[14,23] both field-like torque induced effective field $H_{FL}$ and current-induced Oersted field $H_{Oe}$ contribute to the anti-symmetric component $A$, and the latter dominates it. In contrast, the symmetric component S mainly arises from the damping-like torque induced effective field $H_{DL}$. $S$ and $A$ are the magnitudes of the symmetric and anti-symmetric component of $V_{mix}$ mentioned above, proportional to $H_{DL}$ and $H_{FL}$, respectively. Similar to the method commonly used in such Ta/Py system,[14] here, we also assume that the anti-symmetric component is caused mainly by Oersted field $H_{Oe}$, and the symmetric component dominated by $H_{DL}$. One should note that the contribution to the symmetric component from ISHE due to the spin pumping effect is neglectable for our micro-sized samples. Thus, the SOT efficiency, also called the effective spin Hall angle $\theta_{DL}$, of the studied samples, can be quantitively estimated by the lineshape analysis as following:

$$\theta_{DL} = \frac{S}{A}\frac{e\mu_0 M_S t_N d_F}{\hbar}\left[1 + (4\pi M_{eff}/H_{FMR})\right]^{1/2}, \quad (2)$$

where $M_S$, $t_N$, $d_F$, $A$ and $S$ are the saturation magnetization of Ta(O)(8)/Py(6) films obtained by superconducting quantum interference devices (SQUID), the thickness of Ta and ferromagnetic Py layer, the symmetric and anti-symmetric component of $V_{mix}$, respectively.[4,23] Fig. 2 (d) shows the obtained SOT efficiency $\theta_{DL}$ of all studied samples. The results are substantially consistent with each other for several measurements with different frequencies. The calculated $\theta_{DL}$ of Ta/Py ($Q = 0$) is -0.18, which is very close to the previous reports.[14] Nevertheless, a surprising result is that $\theta_{DL}$ of Ta(O) has a 65% enhancement compare to pure Ta, and reaches to -0.3 when the growth oxygen flow rate $Q$ increases to 0.3% (oxygen concentration n ~ 7.81%).

One may expect that the above lineshape analysis method will lead to some deviation of the estimated $\theta_{DL}$ because the inverse SHE voltage due to spin pumping will be added to the symmetric component $V_s$ and the field-like torque $H_{FL}$ will also generate some anti-symmetric signal $V_a$ [23,25,29]. To accurately estimate the damping-like torque efficiency $\theta_{DL}$, we further studied the dependence of the ST-FMR linewidth on current. The effective damping-like torque efficiency $\theta_{DL}$ can be accurately extracted through dc current-dependent linewidth $\Delta H$ measurements by the following formula:[24,30]

$$\theta_{DL} = \frac{\delta\Delta H/\delta I_{DC}}{\frac{2\pi f}{\gamma}\frac{\sin\theta}{(H_{ext}+2\pi M_{eff})\mu_0 M_S d_F}\frac{\hbar}{2e}}\frac{R_{Py}+R_{Ta(O)}}{R_{Py}}A_C, \quad (3)$$

where $R_{Py}$ and $R_{Ta(O)}$ are the resistance of the ferromagnetic Py layer and Ta(O) layer, respectively, and $A_C$ is the cross-sectional area of Ta(O)/Py bilayer. Fig. 3(a) shows current-dependent linewidth $\Delta H$ at two angles θ = 30° and 210° for four samples with different $Q$. One can easily see that the sample ($Q = 0.3$%) has the maximum slope of linewidth $\delta\Delta H/\delta I_{DC}$. Based on Eq. (3), to quantitatively obtain $\theta_{DL}$, we need to know the resistivity of Ta(O) and Py layers except for $\delta\Delta H/\delta I_{DC}$. Therefore, we separately measured the resistivity of 8 nm Ta(O) layer with different oxygen concentrations ($Q = 0 - 0.53\%$), shown in Fig. 3(b). The resistivity of Ta(O) layer shows a slight increase with increasing the growth oxygen flow rate $Q$, consistent with the low-oxidation level obtained from the above XPS analysis in Fig.1(c). The obtained effective damping-like torque efficiency $\theta_{DL}$ of the studied sample with different $Q$ was summarized in Fig. 3(c). The current-modulation linewidth method shows that Ta(O)(8)/Py(6) ($Q = 0.3\%$) has the maximum $\theta_{DL} = -0.28$, and Ta(8)/Py(6) is -0.17, well consisting with -0.3 ($Q=0.3\%$) and -0.18 ($Q = 0$) obtained by the self-calibrated lineshape analysis above.[23,24]

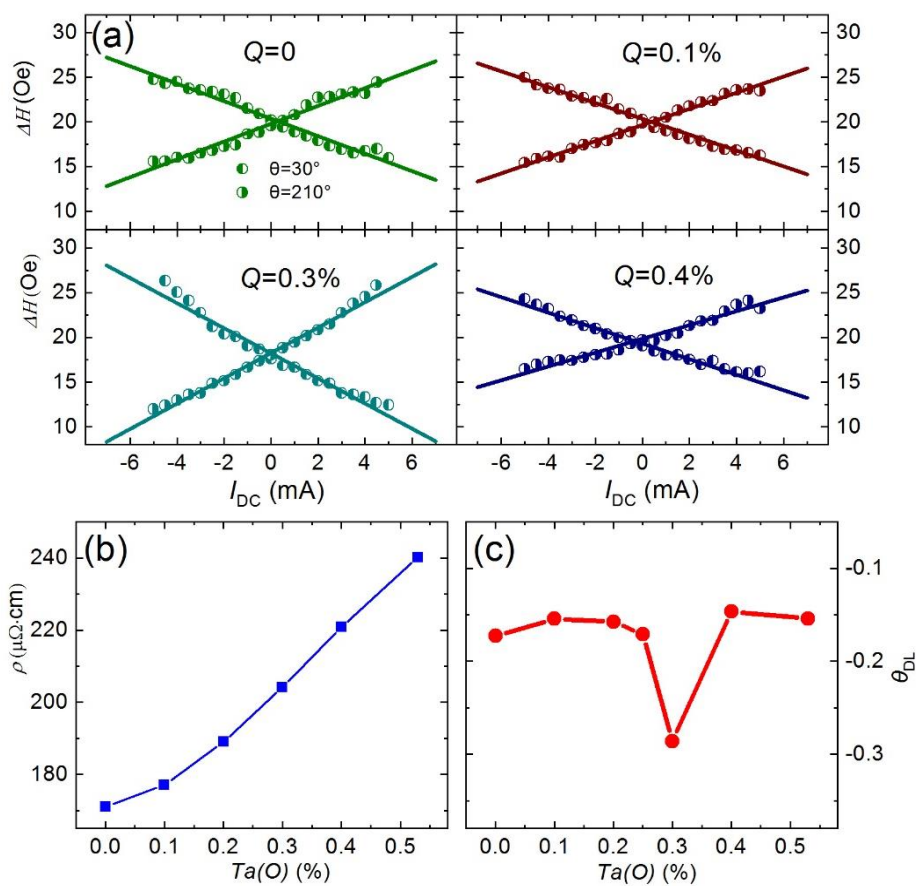

Fig. 3(a) Dependence of linewidth $\Delta H$ on DC current $I_{DC}$ for four different samples ($Q = 0$, 0.1%, 0.3%, 0.4%) at the excitation frequency $f = 4$ GHz. (b) The resistivity of 8

nm thickness Ta(O) films deposited at different oxygen flow rates $Q$. (c) The obtained effective damping-like torque efficiency $\theta_{DL}$ of all studied samples.

As mentioned above, in such Ta(O)/Py system, except for the bulk SHE in Ta(O), the interfacial Rashba-Edelstein effect at the interface between Ta(O) and Py also allows charge-to-spin conversion and exerts a comparable SOT. To get insight on the origin of SOT enhancement in Ta(O)(8)/Py(6) by partially oxidizing strong SHE material Ta layer, we need to experimentally distinguish the contribution of bulk SHE and the interfacial Rashba effect. Therefore, the same measurements with two methods above were performed for Ta(8)/Cu(0, 0.5, 1, 2)/Py(6)/MgO(2) samples. Fig. 4(a) shows the normalized ST-FMR voltage spectra $V_{mix}$ at $f = 4$ GHz. $V_s/V_a$ significantly drops to 0.05 from 0.53 with inserting the 2 nm Cu spacer because of the shunting by low-resistivity Cu spacer. Our results of gradually saturated damping $\alpha$ with increasing Cu thickness, consistent with other reports,[29] indicate that 1 nm thickness Cu spacer layer is enough to eliminate the interface effect of the Ta/Py bilayer due to element mixture and film roughness. Ta(O)(8)/Cu(1)/Py(6) with $Q = 0$ and 0.3% almost have an identical normalized ST-FMR spectra voltage $V_{mix}$ at three different excitation frequencies, shown in Fig. 4(b), which directly suggests that they have the same SOT efficiency and damping constant. Based on the lineshape analysis using Eq. (2), the calculated effective $\theta_{DL}$ is about -0.05 for Ta(8)/Cu(1)/Py(6) due to the shunting of the Cu layer, much lower than the value of -0.18 above-observed in Ta/Py system. Furthermore, through quantifying the shunting effect of the Cu spacer layer based on the parallel-resistor model and long spin diffusion length of Cu,[29,31,32] the bulk SHE-induced $\theta_{DL}$ is calculated to be -0.10 ± 0.01 and -0.11 ± 0.01 (-0.05·$(R_{Cu}+R_{Ta(O)})/R_{Cu}$) for Ta(O) with $Q = 0$ and 0.3%, respectively.

Besides, the bulk SHE-induced $\theta_{DL}$ also can be estimated by analyzing the current-modulation linewidth of FMR spectra. Fig. 4(c) shows that the linewidth of FMR exhibits a linear dependence on the applied current $I_{DC}$ for Ta(O)/Cu(1)/Py ($Q = 0$ and 0.3%). The slope $\delta\Delta H/\delta I_{DC}$ does not show a distinct difference between $Q = 0$ and 0.3% in these inserting Cu samples, but is significantly smaller than that of Ta(O)/Py due to the shutting of the inserting Cu layer. The calculated bulk SHE-induced $\theta_{DL}$ using Eq. (3) are -0.09 ± 0.01 and -0.10 ± 0.01 for Ta(O) with $Q = 0$ and 0.3%, respectively, nearly close to -0.10 ± 0.01 and -0.11 ± 0.01 obtained by the lineshape analysis method above. One should note that the $R_{Py}$ of Eq. (3) needs to be displaced by $R_{Cu/Py} = R_{Cu} \cdot R_{Py}/(R_{Cu}+R_{Py})$. Thus, based on the substantial control experiments on Ta(O)/Py and Ta(O)/Cu/Py, the obtained interfacial Rashba mechanism-induced $\theta_{DL}$ in Ta(O)/Py with $Q = 0$ and 0.3% is -0.08 ± 0.01 and -0.18 ± 0.01, respectively. The respective SOT efficiency $\theta_{DL}$, originated from bulk SHE of Ta(O) and interfacial Rashba mechanisms, was summarized in Fig 4(d). Comparing to Ta/Py, the interfacial Rashba-induced SOT efficiency of Ta(O)/Py ($Q = 0.3\%$) has more than twofold enhancement, even larger than the bulk SHE part of -0.10, while almost no change for the bulk SHE component. Although the SOC of O-2p states is weakly, they could dramatically modify the hybridization of the Ni-3d, Fe-3d, and Ta-5d orbitals near the interface, leading to the change of the wave-function asymmetry near the nucleus. This oxygen-induced wave-function asymmetry can result in the enhancement of the Rashba effects.[20,33] Thus, the reason for SOT enhancement of Ta(O)/Py bilayer is related to the modulation of interfacial Rashba-type SOT due to oxygen-induced orbital hybridization cross the interface.

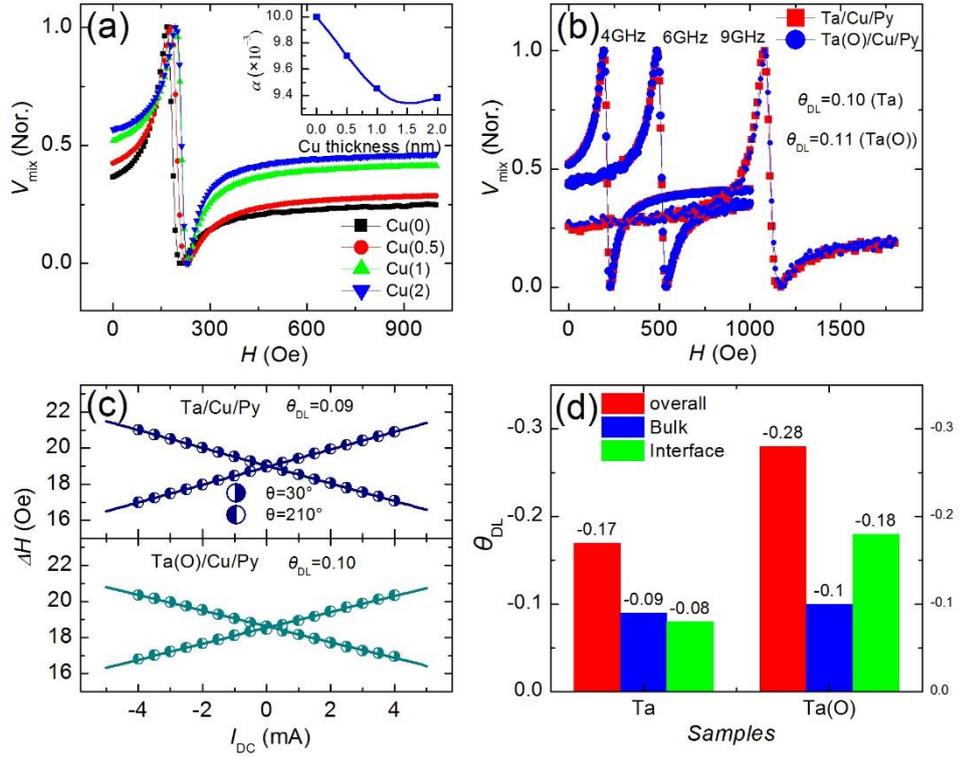

Fig. 4(a) Normalized ST-FMR mixing voltage $V_{mix}$ spectra of Ta(8)/Cu(0, 0.5, 1 and 2)/Py(6)MgO(2) at excitation frequency $f = 4$ GHz. The inset shows Cu thickness dependence of the damping α for Ta/Cu/Py samples. (b) Normalized $V_{mix}$ spectra of Ta(O)(8)/Cu(1)/Py(6)/MgO(2) with $Q = 0$ and 0.3% at three $f = 4$ GHz, 6 GHz and 9 GHz. (c) Dependence of linewidth $\Delta H$ on DC current $I_{DC}$ of Ta(O)(8)/Cu(1)/Py(6) with $Q = 0$ and 3% at $f = 4$ GHz and θ = 30º and 210º. (d) The respective SOT efficiency $\theta_{DL}$ originated from the bulk spin Hall effect and the interfacial Rashba effect.

In summary, we find that the SOT efficiency $\theta_{DL}$ of Ta(O)/Py bilayers can be modulated by fine-tuning of oxygen in heavy metal Ta film and reach to the maximum charge-spin conversion rate of ~ -0.3 at the oxygen concentration of 7.8% from ~ -0.18 in the pure Ta via two methods of symmetrical analysis of ST-FMR and current-modulated damping constant. We distinguish the contribution rates of the bulk SHE and of interface Rashba effect by inserting the copper spacer layer between Ta(O) and Py, and demonstrate that the enhancement of $\theta_{DL}$ is related to the modulation of interfacial Rashba-Edelstein effect via oxygen-induced orbital hybridization at Ta(O)/Py interface. More importantly, the interface Rashba-SOT efficiency has more than a twofold increase, and even become more significant than the bulk SHE mechanism. Our results indicate that modulation of interfacial coupling via oxygen-induced orbital hybridization can be an alternative approach and method to develop more energy-efficient spintronic devices.

We acknowledge support from the National Key Research and Development Program of China (2016YFA0300803), the National Natural Science Foundation of China (No.11774150, No. 12074178, No. 12004171), the Applied Basic Research Programs of Science and Technology Commission Foundation of Jiangsu Province (Grant No. BK20170627), and the Open Research Fund of Jiangsu Provincial Key Laboratory for Nanotechnology.

1. A.G. M. Jansen M. Tsoi, W. -C. Chiang, M. Seek, V. Tsoi, and P.Wyder, Phys. Rev. Lett. **80** (1998).
2. D. C. Ralph E. B. Myers, J. A. Katine, R. N. Louie, R. A. Buhrman, Science **285,** (1999).
3. J.C. Slonczewski, J. Magn. Magn. Mater. **159 L1** (1996); J. Magn. Magn. Mater. **195**, L261 (1996).
4. L. Liu, T. Moriyama, D. C. Ralph, and R. A. Buhrman, Phys. Rev. Lett. **106** (3), 036601 (2011).
5. K. Kondou, R. Yoshimi, A. Tsukazaki, Y. Fukuma, J. Matsuno, K. S. Takahashi, M. Kawasaki, Y. Tokura, and Y. Otani, Nat. Phys. **12** (11), 1027 (2016).
6. Jairo Sinova, Sergio O. Valenzuela, J. Wunderlich, C. H Back, and T. Jungwirth, Rev. Mod. Phys. **87** (4), 1213 (2015).
7. Tomas Jungwirth, Jörg Wunderlich, and Kamil Olejník, Nat. Mater. **11** (5), 382 (2012).
8. I. M. Miron, G. Gaudin, S. Auffret, B. Rodmacq, A. Schuhl, S. Pizzini, J. Vogel, and P. Gambardella, Nat. Mater. **9** (3), 230 (2010).
9. Q. Shao, G. Yu, Y. W. Lan, Y. Shi, M. Y. Li, C. Zheng, X. Zhu, L. J. Li, P. K. Amiri, and K. L. Wang, Nano lett. **16** (12), 7514 (2016).
10. P. Noel, F. Trier, L. M. Vicente Arche, J. Brehin, D. C. Vaz, V. Garcia, S. Fusil, A. Barthelemy, L. Vila, M. Bibes, and J. P. Attane, Nature **580** (7804), 483 (2020).
11. J. C. Sanchez, L. Vila, G. Desfonds, S. Gambarelli, J. P. Attane, J. M. De Teresa, C. Magen, and A. Fert, Nat. Commun. **4**, 2944 (2013).
12. I. M. Miron, K. Garello, G. Gaudin, P. J. Zermatten, M. V. Costache, S. Auffret, S. Bandiera, B. Rodmacq, A. Schuhl, and P. Gambardella, Nature **476** (7359), 189 (2011).

13   M. Baumgartner, K. Garello, J. Mendil, C. O. Avci, E. Grimaldi, C. Murer, J. Feng, M. Gabureac, C. Stamm, Y. Acremann, S. Finizio, S. Wintz, J. Raabe, and P. Gambardella, Nat. Nanotechnol. **12** (10), 980 (2017).
14   Luqiao Liu, Chi-Feng Pai, Y. Li, H. W. Tseng, D. C. Ralph, R. A. Buhrman, Science **336**, 555 (2012).
15   Chi-Feng Pai, Luqiao Liu, Y. Li, H. W. Tseng, D. C. Ralph, and R. A. Buhrman, Appl. Phys. Lett. **101** (12) 122404 (2012).
16   Lijun Ni, Zhendong Chen, Xianyang Lu, Yu Yan, Lichuan Jin, Jian Zhou, Wencheng Yue, Zhe Zhang, Longlong Zhang, Wenqiang Wang, Yong-Lei Wang, Xuezhong Ruan, Wenqing Liu, Liang He, Rong Zhang, Huaiwu Zhang, Bo Liu, Ronghua Liu, Hao Meng, and Yongbing Xu, Appl. Phys. Lett. **117** (11) 112402 (2020).
17   A. Soumyanarayanan, N. Reyren, A. Fert, and C. Panagopoulos, Nature **539** (7630), 509 (2016).
18   A. Manchon, H. C. Koo, J. Nitta, S. M. Frolov, and R. A. Duine, Nat. Mater. **14** (9), 871 (2015).
19   Yu. A. Bychkov and E. I. Rashba, JETP Lett. **39** (2), 020078 (1984).
20   O. Krupin, G. Bihlmayer, K. Starke, S. Gorovikov, J. E. Prieto, K. Döbrich, S. Blügel, and G. Kaindl, Phys. Rev. B **71** (20) 201403 (2005).
21   R. H. Liu, W. L. Lim, and S. Urazhdin, Phys. Rev. B **89** (22) 220409 (2014).
22   H. Yang, G. Chen, A. A. C. Cotta, A. T. N'Diaye, S. A. Nikolaev, E. A. Soares, W. A. A. Macedo, K. Liu, A. K. Schmid, A. Fert, and M. Chshiev, Nat. Mater. **17** (7), 605 (2018).
23   Takeo Ohno Hongyu An, Yusuke Kanno, Yuito Kageyama, Yasuaki Monnai, Hideyuki Maki, Ji Shi, Kazuya Ando, Sci. Adv. **4** eaar2250 (2018).
24   K. U. Demasius, T. Phung, W. Zhang, B. P. Hughes, S. H. Yang, A. Kellock, W. Han, A. Pushp, and S. S. P. Parkin, Nat. Commun. **7**, 10644 (2016).
25   Yi Wang, Rajagopalan Ramaswamy, and Hyunsoo Yang, J. Phys. D: Appl. Phys. **51** (27) 273002 (2018).
26   Michael Harder, Yongsheng Gui, and Can-Ming Hu, Physics Reports **661**, 1 (2016).
27   L. Bai, P. Hyde, Y. S. Gui, C. M. Hu, V. Vlaminck, J. E. Pearson, S. D. Bader, and A. Hoffmann, Phys. Rev. Lett. **111** (21), 217602 (2013).
28   Y. Tserkovnyak, A. Brataas, and G. E. Bauer, Phys. Rev. Lett. **88** (11), 117601 (2002).
29   Xiao Wang Jing Zhou, Yaohua Liu, Jihang Yu, Huixia Fu, Liang Liu, Shaohai Chen, Jinyu Deng, Weinan Lin, Xinyu Shu, Herng Yau Yoon, Tao Hong, Masaaki Matsuda, Ping Yang, Stefan Adams, Binghai Yan, Xiufeng Han, Jingsheng Chen, Sci. Adv. **5** eaau6696 (2019).
30   K. Ando, S. Takahashi, K. Harii, K. Sasage, J. Ieda, S. Maekawa, and E. Saitoh, Phys. Rev. Lett. **101** (3), 036601 (2008).
31   H. An, Y. Kageyama, Y. Kanno, N. Enishi, and K. Ando, Nat Commun **7**, 13069 (2016).
32   T. Gao, A. Qaiumzadeh, H. An, A. Musha, Y. Kageyama, J. Shi, and K. Ando, Phys. Rev. Lett. **121** (1), 017202 (2018).
33   Yuya Tazaki Yuito Kageyama, Hongyu An, Takashi Harumoto, Tenghua Gao, Ji Shi, kazuya Ando, Sci. Adv. **5** eaax4278 (2019).